\documentclass[aps,pre,twocolumn,showpacs,10pt]{revtex4-1}
\usepackage{amsfonts}
\usepackage{amsmath}
\usepackage{amssymb}
\usepackage{graphicx}
\usepackage{mathptmx}
\usepackage{helvet}

\begin{document}

\title{The ponderomotive force due to the intrinsic spin in extended fluid
and kinetic models}

\author{Martin Stefan}

\email{martin.stefan@physics.umu.se}

\affiliation{Department of Physics, Ume{\aa} University, SE--901 87 Ume{\aa},
Sweden}

\author{Jens Zamanian}

\email{jens.zamanian@physics.umu.se}

\affiliation{Department of Physics, Ume{\aa} University, SE--901 87 Ume{\aa},
Sweden}

\author{Gert Brodin}

\email{gert.brodin@physics.umu.se}

\affiliation{Department of Physics, Ume{\aa} University, SE--901 87 Ume{\aa},
Sweden}

\author{Amar Prasad Misra}

\email{apmisra@visva-bharati.ac.in}

\affiliation{Department of Physics, Ume{\aa} University, SE--901 87 Ume{\aa},
Sweden}

\author{Mattias Marklund}

\email{mattias.marklund@physics.umu.se}

\affiliation{Department of Physics, Ume{\aa} University, SE--901 87 Ume{\aa},
Sweden}

\date{02 December, 2010}

\begin{abstract}
In this paper we calculate the contribution to the ponderomotive force
in a plasma from the electron spin using a recently developed model.
The spin-fluid model used in the present paper contains spin-velocity
correlations, in contrast to previous models used for the same purpose.
Is its then found that previous terms for the spin-ponderomotive force
are recovered, but also that additional terms appear. Furthermore,
the results due to the spin-velocity correlations are confirmed using
the spin-kinetic theory. The significance of our results is discussed. 
\end{abstract}

\pacs{52.25.Dg, 52.25.Xz, 52.35.Hr, 52.35.Mw}

\maketitle

\section{Introduction}

The ponderomotive force plays a crucial role in the nonlinear dynamics
of plasmas. Phenomena induced by the ponderomotive force include,
e.g. wakefield generation \cite{gorbunov87,bingham2007}, soliton
formation, self-focusing and wave collapse \cite{berge98}. The classical
expression for the ponderomotive force in a magnetized plasma first
derived by Karpman \emph{et al} \cite{Karpman-pond} was recently
generalized to account for quantum mechanical effects, in particular
due to the electron spin \cite{brodin2010}. The physics of quantum
plasmas (for recent reviews, see e.g. \cite{Shukla-Eliasson,Manfredi2005})
has recently received much interest due to applications in, e.g.
quantum wells \cite{Manfredi-quantum-well}, plasmonics \cite{Atwater-Plasmonics}
and spintronics \cite{Spintronics}. In particular, the dynamical
effects due to the intrinsic spin of the electron has been investigated
using both fluid \cite{marklund-brodin,brodin2008,stefan2010} and semi-classical kinetic
approaches \cite{cowley,kulsrud,balescu1988,zhang,brodin-etal}. In
Ref.\ \cite{brodin2010} the ponderomotive force related to the magnetic
dipole moment of the electrons (due to the spin) was shown to induce
a spin-polarized plasma, i.e.\ the spin-up and spin-down states of
the electrons are separated. This effect was shown to be pronounced
also for a plasma of relatively modest density \cite{brodin2010},
and the expressions has recently been applied to astrophysical plasmas
\cite{AP-Misra2010a,AP-Misra2010b}. Furthermore, the spin-polarization,
in turn, was shown to induce cubically nonlinear terms that may influence
the high-frequency dynamics \cite{brodin2010}.

When the spin contribution of electrons to the ponderomotive force was calculated in Ref.\ \cite{brodin2010}, a relative simple fluid model was used. 
The model included the magnetic dipole force, spin-precession as well
as the magnetization current due to the spin, but not the spin-velocity
correlations. However, in a recent work \cite{Zamanian2010b} a fluid
model of spin that includes spin-velocity correlations was shown to
capture the spin effects of kinetic theory much more accurately, although
the comparison between fluid and kinetic theory in Ref. \cite{Zamanian2010b} was 
limited to linear phenomena.

In the present paper, we aim to improve the calculation made in Ref.\
\cite{brodin2010} to capture the effects of the spin-velocity tensor
using the newly developed four-equation hierarchy \cite{Zamanian2010b}
for spin-fluid dynamics. Furthermore, in order to validate this model,
we compare our results with the full kinetic treatment from the equations
derived in Ref.\ \cite{zamanian2010a}. The set-up is that of a weakly nonlinear
electromagnetic pulse propagating parallelly to an external magnetic
field. It is then confirmed that a spin-ponderomotive term of the
same kind as in previous calculations exists (together with the classical
ponderomotive force), but it is also found that the spin-velocity
correlations induce an additional term. This new term, due to the
spin-velocity correlations, gives rise to a force in the same directions
for all particles, regardless of the spin-states (up or down) relative
to the external magnetic field. This is in contrast to the previously
found term \cite{brodin2010} which acted in opposite directions for
spin-up and down states. Furthermore, the new term has higher order
resonances at the spin precession frequency, as compared to the previous
term \cite{brodin2010}. The comparisons with kinetic theory give
a perfect agreement in the low-temperature limit. This comparison
also includes a calculation of the low-frequency magnetization induced
by the spin-ponderomotive force. The main purpose of the present paper
is to validate the newly developed model against kinetic theory also
in the regime of nonlinear perturbations. However, we also point out
that the newly found contribution to the spin-ponderomotive force
is likely to be important for the nonlinear evolution in quantum plasmas.

\section{Fluid description}

In Ref.\ \cite{Zamanian2010b} (see also Ref.\ \cite{haas2010}) a fluid moment hierarchy was derived
from a quantum kinetic theory with spin, extending previous hydrodynamical
models for particles with spin-$1/2$. The time evolution equations
are given by \begin{align}
\partial_{t}n+\partial_{i}(nv_{i})={} & 0\label{Cont}\\
m(\partial_{t}+v_{j}\partial_{j})v_{i}={} & q\left(E_{i}+\epsilon_{ijk}v_{j}B_{k}\right)+\frac{1}{n}\partial_{j}P_{ij}\notag\\
 & +\frac{2\mu}{\hbar}S_{j}\partial_{i}B_{j}\\
(\partial_{t}+v_{j}\partial_{j})S_{i}={} & -\frac{2\mu}{\hbar}\epsilon_{ijk}B_{j}S_{k}-\frac{1}{nm}\partial_{j}\Sigma_{ij}\label{S}\end{align}
\begin{align}
(\partial_{t}+v_{k}\partial_{k})P_{ij} & =-P_{ik}\partial_{k}u_{j}-P_{jk}\partial_{k}u_{i}-P_{ij}\partial_{k}u_{k}\notag\\
 & +\frac{q}{m}\epsilon_{imn}P_{jm}B_{n}+\frac{q}{m}\epsilon_{jmn}P_{im}B_{n}\notag\\
 & +\frac{2\mu}{\hbar m}\Sigma_{ik}\partial_{j}B_{k}+\frac{2\mu}{\hbar m}\Sigma_{jk}\partial_{i}B_{k}-\partial_{k}Q_{ijk}\label{Pressure}\\
(\partial_{t}+v_{k}\partial_{k})\Sigma_{ij}={} & -\Sigma_{ij}\partial_{k}v_{k}-\Sigma_{ik}\partial_{k}v_{j}-P_{jk}\partial_{k}S_{i}\notag\\
 & +\frac{q}{m}\epsilon_{jkl}\Sigma_{ik}B_{l}+\frac{2\mu}{\hbar}\epsilon_{ikl}\Sigma_{kj}B_{l}\notag\\
 & +\frac{\hbar\mu n}{2}\partial_{j}B_{i}-\frac{2\mu n}{\hbar}S_{i}S_{k}\partial_{j}B_{k},\label{sigma}\end{align}
where $n$ is the particle number density, $\mathbf{v}$ is the fluid
velocity, $q$ is the particle charge, $m$ is the mass, $P_{ij}$
is the pressure tensor, $S_{i}$ is the spin density which is defined
to have length $\hbar/2$ for a coherent spin state, and $Q_{ijk}$
is the heat flux tensor. The magnetic moment of the particle is given
by $\mu=g q \hbar/4m$, where $g$ is the g-factor of the particle in
question ($g=2.0023$ for electrons). Summation over repeated indices
$i,j,k\dots=1,2,3$ is understood and we have used the notation $\partial_{t}=\partial/\partial t$
and $\partial_{i}=\partial/\partial x_{i}$ and $\epsilon_{ijk}$
is the antisymmetric Levi-Civita tensor. The tensor $\Sigma_{ij}$
is the spin-velocity tensor which correlates spin and velocity and
captures, in a macroscopic description, how different forces acts on
different parts of the spin probability distribution. In Ref.\ \cite{Zamanian2010b}
inclusion of $\Sigma_{ij}$ was shown to reproduce some of the more
subtle effects predicted by kinetic theory in the linear regime. Note that
quantum mechanical effects associated with particle dispersion (e.g. the Bohm potential) could have been
included in the above model. However, since such effects will not influence the ponderomotive force of transverse
waves propagating along a magnetic field \cite{brodin2010}, which is our problem of consideration, we do not include such terms here.

Next, we consider circularly polarized electromagnetic waves propagating
parallel to an external magnetic field, $\mathbf{B}_{0}=B_{0}\hat{z}$,
and use the following ansatz \begin{align}
\mathbf{E} & =\frac{1}{2}\left[\tilde{\mathbf{E}}(z,t)e^{i\left(kz-\omega t\right)}+\tilde{\mathbf{E}}^{\ast}(z,t)e^{-i\left(kz-\omega t\right)}\right],\notag\\
\mathbf{B} & =\frac{1}{2}\left[\tilde{\mathbf{B}}(z,t)e^{i\left(kz-\omega t\right)}+\tilde{\mathbf{B}}^{\ast}(z,t)e^{-i\left(kz-\omega t\right)}\right],\label{Ansatz-1}\end{align}
where the amplitudes are assumed to vary much slower than the exponential
factors, and the star denotes complex conjugates. Since the basic
wave modes propagating parallel to $\mathbf{B}_{0}$ are either left-
or right-circularly polarized, we have $\tilde{\mathbf{E}},\tilde{\mathbf{B}}\propto$
$\widehat{\mathbf{x}}\pm i\widehat{\mathbf{y}}$. Furthermore, all
perturbations are small, such that weakly nonlinear perturbation theory
is applicable. The equilibrium density and spin density will be denoted
by $n_{0}$ and $S_{0}$, respectively.

\subsection{Spin-Ponderomotive Force}

When thermal effects are small, the classical (superscipt $\mathrm{cl}$)
low-frequency (subscript ($\mathrm{lf}$)) ponderomotive force corresponding
to the ansatz (\ref{Ansatz-1}) (with transverse fields) is 
\begin{equation}
F_{z\mathrm{lf}}^{\mathrm{cl}}=-\frac{q^{2}}{2m\omega(\omega\mp\omega_{c})}\left[\partial_{z}\mp\frac{k\omega_{c}}{\omega(\omega\mp\omega_{c})}\partial_{t}\right]|E_{\pm}|^{2}.\label{Karpman}
\end{equation}
where $\omega_{c}\equiv qB_{0}/m$ is the electron cyclotron frequency
\cite{Cyclotron-note} (We will here limit ourselves to electrons,
although all results in Sec. II can straightforwardly be generalized
to any particle species fulfilling Eqs. (\ref{Cont})-(\ref{sigma})).
Equation (\ref{Karpman}) was first derived by Karpman \emph{et al}
\cite{Karpman-pond}, and has been verified by many subsequent authors.
On the other hand, the spin-dependent part of the ponderomotive force
was recently calculated by Brodin \emph{et al} \cite{brodin2010},
but starting from a model without the spin-velocity tensor. Now, we
follow their outline to calculate the spin contribution to the ponderomotive
force, but taking also the parts originating from $\Sigma_{ij}$ into
account.

To find the spin dependent part of the ponderomotive force, we first
linearize the Eq.\ \eqref{sigma} and note that the only components
of the spin-velocity vector that have a driving term are $\Sigma_{13}$
and $\Sigma_{23}$ and thus we define $\Sigma_{\pm}=\tilde{\Sigma}_{13}\pm i\tilde{\Sigma}_{23},$
$B_{\pm}=\tilde{B}_{x}\pm i\tilde{B}_{y}$ and $S_{\pm}=S_{x}\pm iS_{y}$.
Neglecting the slow derivatives, i.e.\ derivatives acting on the
amplitudes, we get the linear solution to Eq.\ \eqref{sigma} given
by 
\begin{equation}
\Sigma_{\pm}=-\frac{\hbar\mu n_{0}k}{4(\omega\mp\omega_{cg})}B_{\pm},
\label{Sigma_1}
\end{equation}
where $\omega_{cg}\equiv2\mu B_{0}/\hbar$ is the spin-precession
frequency (which is close to the cyclotron frequency for electrons,
as $\left\vert g\right\vert =2.0023$). Iterating by plugging this
back into Eq.\ \eqref{sigma} we find the correction due to the slow
derivatives as \begin{equation}
\Sigma_{\pm}=-\frac{\hbar\mu n_{0}}{4(\omega\mp\omega_{cg})}\left(k-\frac{ik}{\omega\mp\omega_{cg}}\partial_{t}-i\partial_{z}\right)B_{\pm}.\label{Sigma_pm}\end{equation}
This last step could also be done by simply making the substitutions
$\omega\rightarrow\omega+i\partial_{t}$ and $k\rightarrow k-i\partial_{z}$
in Eq.\ \eqref{Sigma_1} and expanding to lowest order in the slow
derivatives. From Faraday's law the electric and magnetic fields are
similarly related by 
\begin{equation}
B_{\pm}=\pm i\frac{k}{\omega}E_{\pm}\pm\frac{1}{\omega}\frac{\partial E_{\pm}}{\partial z}\pm\frac{k}{\omega^{2}}\frac{\partial E_{\pm}}{\partial t}\label{xtra}
\end{equation}
 The lowest order approximation, $B_{\pm}=\pm i(k/\omega)E_{\pm}$,
can be used to switch between magnetic and electric fields in, e.g.
the right-hand side of Eq.\ (\ref{Karpman}). Repeating the above steps
for Eq.\ \eqref{S} and using Eq.\ \eqref{Sigma_pm} we can express
the spin variable as 
\begin{align}
S_{\pm} & =\mp\frac{\mu S_{0}}{\hbar(\omega\mp\omega_{cg})}B_{\pm}+\frac{\hbar\mu}{4m(\omega\mp\omega_{cg})^{2}}\nonumber \\
 & \times\left[-k^{2}+i\left(\pm\frac{4mS_{0}}{\hbar^{2}}+\frac{2k^{2}}{\omega\mp\omega_{cg}}\right)\partial_{t}+2ik\partial_{z}\right]B_{\pm}.\label{S_pm}
 \end{align}
 The spin-ponderomotive force density is then given by the last term
of Eq.\ \eqref{S} and can be written as \begin{equation}
F_{z\mathrm{lf}}=\frac{\mu}{2\hbar}\left(S_{\pm}\partial_{z}B_{\pm}^{\ast}+S_{\pm}^{\ast}\partial_{z}B_{\pm}\right).\end{equation}
It should be noted that the pressure tensor in Eq.\ \eqref{Pressure}
actually has quadratically nonlinear source terms proportional to
the magnetic moment which, in principle, could contribute to a low-frequency
spin force, in addition to that originating directly from the magnetic
dipole force. However, following the same calculation scheme as outlined
above, it can be verified that these terms do not contribute to the
leading order in the slow derivative expansion. Thus, using Eq.\ \eqref{S_pm}
to the first order in slow derivatives we obtain 
\begin{align}
F_{z\mathrm{lf}}= & \mp\frac{\mu^{2}S_{0}}{2\hbar^{2}(\omega\mp\omega_{cg})}\left(\partial_{z}-\frac{k}{\omega\mp\omega_{cg}}\partial_{t}\right)\left\vert B_{\pm}\right\vert ^{2}\notag\\
 & {}+\frac{\mu^{2}k^{2}}{8m(\omega\mp\omega_{cg})^{2}}\left[\partial_{z}+\frac{2k}{\omega\mp\omega_{cg}}\partial_{t}\right]\left\vert B_{\pm}\right\vert ^{2}.\label{fluid-ponderomotive}\end{align}
Comparing this force with the result of Ref.\ \cite{brodin2010}
we see that the first terms (proportional to $S_{0}$) corresponds
exactly to their result {[}after a correction of factor $2$ in Eq.\ (13) 
of Ref.\ \cite{brodin2010} has been made]. The second term
comes in due to the inclusion of the spin-velocity tensor (which was
neglected in Ref.\ \cite{brodin2010}). It has an extra factor $\hbar k^{2}/2m\left(\omega\mp\omega_{cg}\right)$
compared to the first term and could in a sense be viewed as a higher-order
quantum correction. This term might, however, dominate over the first
term in some cases where the zeroth order magnetization is small.
For thermal equilibrium we have $S_{0}=(\hbar/2)\tanh\left(\mu B_{0}/k_{B}T\right)$
where $\tanh\left(\mu B_{0}/k_{B}T\right)$ $\simeq\mu B_{0}/k_{B}T$
for a moderate magnetic field strength, and thus both the terms of
the spin-ponderomotive force are quadratic in $\hbar$. In this case,
the ratio of the first to second term scales as $\sim m\omega_{cg}\left(\omega-\omega_{cg}\right)/k_{B}Tk^{2}$,
which could be both smaller or larger than unity. On the other hand,
for a higher magnetic field strength for which $\tanh\left(\mu B_{0}/k_{B}T\right)\sim1$,
the same ratio is scaled as $\sim m\left(\omega-\omega_{cg}\right)/\hbar k^{2},$
which can also be either smaller or larger than the unity. Thus, depending
on the parameter regimes we consider for a specific problem,
both the classical and the spin-ponderomotive force may either be
comparable or even dominate over each other \cite{AP-Misra2010a,AP-Misra2010b}. 
As for example, for a wave frequency close to the electron-cyclotron frequency, the spin contribution to 
the ponderomotive force can, indeed, dominate over the classical one when $\hbar k^2/m\omega\gg1$ \cite{brodin2010,AP-Misra2010a,AP-Misra2010b}.  

A slightly different perspective was taken in, e.g. Ref. \cite{brodin2010}.
There a two-fluid model for electrons was used, where spin-up and down
states were treated as different species, and thus the two species
have $S_{0}=\pm(\hbar/2)$. As a consequence, the force on each separate
species {[}due to the first term of Eq. (\ref{fluid-ponderomotive}){]}
is typically stronger {[}since the diminishing factor $\mu B_{0}/k_{B}T$
disappears from the first terms, the ratio between the two types of
terms are now $\hbar k^{2}/2m\left(\omega\mp\omega_{cg}\right)${]},
but on the other hand, the forces on the two spin-populations act
in opposite directions due to the terms proportional to $S_{0}$.
Since the physics is different depending on whether the terms induce
spin-polarization or not, both types of terms may be needed for an
accurate description even in the cases when one of the terms is much
larger than the other. As argued in Ref. \cite{Zamanian2010b}, the
two-fluid model captures some of the effects of a kinetic theory,
but it is less needed when the spin-velocity correlations of the $\Sigma_{ij}$-tensor
is included in the model. We will return to this issue of one-fluid
versus two-fluid models in a little more detail below. As a final
note of this subsection we point out that for slowly modulated waves
as considered here we may in many cases use the approximate relation
$\partial_{t} \approx-v_{g}\partial_{z}$ (where $v_g$ is the group velocity) 
to simplify Eqs.\ (\ref{Karpman})
and (\ref{fluid-ponderomotive}), although more accurate expressions
might be needed in case the terms from the spatial and temporal derivative
tend to cancel.


\subsection{Low-frequency magnetization}

The slowly modulated wave also gives rise to a nonlinearly induced
low-frequency magnetization. To further investigate the fluid model
we next calculate the low frequency magnetization defined as 
\begin{equation}
\mathbf{M}_{\mathrm{lf}}=\mu\left(n_{0}\mathbf{S}_{\mathrm{lf}}+n_{\mathrm{lf}}\mathbf{S}_{0}\right).\label{mdef}
\end{equation}
We are particularly interested in the case where the low-frequency
magnetization is induced in the absence of prior magnetization, and
hence we take $\mathbf{S}_{0}=0$ for simplicity. As $S_{0}=(\hbar/2)\tanh\left(\mu B_{0}/k_{B}T\right)$
in a thermal equilibrium background, this may be a useful approximation
also from a practical perspective, provided $\mu B_{0}/k_{B}T\ll1$.

Applying Eq. (\ref{S}), considering the low-frequency time scale
and keeping terms up to the second order in amplitude, we obtain \begin{equation}
\partial_{t}S_{\mathrm{lf}}^{i}=\mp\frac{\mu}{2\hbar}\left(iB_{\pm}S_{\pm}^{\ast}+\mathrm{c.c}.\right)\delta_{i3}-\frac{1}{n_{0}m}\partial_{j}\Sigma_{\mathrm{lf}}^{ij},\end{equation}
where c.c. denotes the complex conjugate. The first term can be calculated
from Eq. (\ref{S_pm}), and we note that only terms involving slow
derivatives survive when adding the complex conjugates. As for the
second term, only the $\Sigma_{\mathrm{lf}}^{i3}$ components need
to be calculated. Thus, out of these three, the only component with
a nonzero driving term is \begin{equation}
\partial_{t}\Sigma_{\mathrm{lf}}^{33}=\pm i\frac{\mu}{2\hbar}\Sigma_{\pm}B_{\pm}^{\ast}+\mathrm{c.c.}\label{sigma33}\end{equation}
where $+\mbox{c.c.}$ means adding the complex conjugate of the right-hand
side. Combining the above formulas with the linear results from the
previous section, Eqs. (\ref{Sigma_pm}) and (\ref{S_pm}) we obtain
\begin{align}
\partial_{t}^{2}M_{\mathrm{lf}}= & \frac{\mu^{3}n_{0}}{4m\hbar(\omega\mp\omega_{cg})}\left(\partial_{z}-\frac{2k}{\omega\mp\omega_{cg}}\partial_{t}\right)\notag\\
 & \times\left(\partial_{z}+\frac{k}{\omega\mp\omega_{cg}}\partial_{t}\right)|B_{\pm}|^{2}.\label{magnetization}\end{align}


\subsection{Magnetization in spin two-fluid models}


In Ref.\ \cite{brodin2010} a similar problem was studied, where the
first term of Eq.\ (\ref{fluid-ponderomotive}) induced opposite density
perturbations for particles with initial spin-up (i.e. with $S_{0}=+\hbar/2$)
and for particles with initial spin-down (with $S_{0}=-\hbar/2)$).
As pointed out above, it is straightforward to include such a formalism
within the present theory, we just consider two electrons fluids which
are identical in all aspects, except that the unperturbed spin in
the $z$-direction is $+\hbar/2$ or $-\hbar/2$. The expression
Eq.\ (\ref{fluid-ponderomotive}) is then the same, except that we
substitute $S_{0}=\pm(\hbar/2)$ where $+$ ($-$) stands for up (down)
species, rather than the thermodynamic equilibrium expression $S_{0}=(\hbar/2)\tanh\left(\mu B_{0}/k_{B}T\right)$
of the one-fluid theory. 

Let us now calculate the two-fluid version corresponding to Eq.\ \eqref{magnetization},
which is still evaluated when $\mu B_{0}/k_{B}T\ll1$ such that
$S_{0}$ of the one-fluid theory is negligible. This corresponds to
equal densities of the initial spin-up and down- populations in the
two-fluid model. Somewhat surprisingly we obtain exact agreement with
Eq.\ \eqref{magnetization} using such a two-fluid theory. However,
contributions to the magnetization that come from the spin-velocity
correlations in the one-fluid model is, to some extent, replaced by
contributions that come from density perturbations of the two-species
in the two-fluid model. Specifically, we note that our current model (which, as we
recall, includes a modified ponderomotive force due to the spin-velocity
tensor) agrees with the density perturbations of the spin-up and spin-down
populations of the less elaborate model used in Ref.\ \cite{brodin2010}.
Furthermore, we stress that the additional spin-ponderomotive force term does not contribute
to the spin-polarization. Nevertheless, the corresponding magnetization found here
can be much lower in certain cases, as compared to the model without
spin-velocity correlations. In particular, this holds for the specific
case of an unmagnetized plasma $(B_{0}=0)$, and a weakly dispersive
driver, i.e. from Eq. \eqref{magnetization} we find that for $\partial_{t}=-v_{g}\partial_{z}$,
$B_{0}=0$ and zero dispersion (i.e. $v_{g}=\omega/k$), the right-hand
side of Eq.\ \eqref{magnetization} vanishes. This may seem surprising,
as intuitively one would expect that a net density difference between
the up- and down-species (as found in \ a two-fluid model both with
and without spin-velocity correlations) implies a net magnetization.
However, the reason of why this does not give rise to a finite magnetization
for the given assumptions specified above, is that the density difference
is compensated by a rotation of the spin direction, which also is
of second order in the transverse wave field amplitudes. 

In a two-fluid model [where the unperturbed spin of the species are
$\mathbf{S}_{0}=\pm(\hbar/2)\widehat{\mathbf{z}}$] the magnetization
for each species is then given by Eq.\ \eqref{mdef} and the total
low-frequency magnetization is obtained as \begin{align}
M_{z\mathrm{lf}}= & {}M_{z\mathrm{lf}\uparrow}+M_{z\mathrm{lf}\downarrow}\notag\\
 & =\frac{{}2\mu}{\hbar}[n_{0}(S_{z\mathrm{lf}\uparrow}-S_{z\mathrm{lf}\downarrow})+(n_{\mathrm{lf}\uparrow}-n_{\mathrm{lf}\downarrow})S_{0}],\end{align}
with $S_{0}=\hbar/2$. For the assumptions specified above where the
induced magnetization vanish, we accordingly have $(S_{z\mathrm{lf}\uparrow}-S_{z\mathrm{lf}\downarrow})=-(n_{\mathrm{lf}\uparrow}-n_{\mathrm{lf}\downarrow})\hbar/2n_{0}$.

\section{Kinetic approach}

In order to further investigate the fluid model of Eqs.\ \eqref{Cont}-\eqref{sigma},
we now consider the same situations as above but using a kinetic model.
The evolution of the scalar distribution for a spin particle is, in
the long scalelength limit, given by \cite{zamanian2010a} 
\begin{align}
 & \partial_{t}f+\mathbf{v}\cdot\nabla_{x}f+\frac{q}{m}\left(\mathbf{E}+\mathbf{v}\times\mathbf{B}\right)\cdot\nabla_{v}f\notag\\
 & +\frac{\mu}{m}\nabla_{x}\left(\mathbf{B}\cdot\mathbf{s}+\mathbf{B}\cdot\nabla_{s}\right)\cdot\nabla_{v}f+\frac{2\mu}{\hbar}(\mathbf{s}\times\mathbf{B})\cdot\nabla_{s}f=0.\label{k1}\end{align}
Here, the distribution function $f=f(\mathbf{x},\mathbf{v},\mathbf{s},t)$
contains a spin variable $\mathbf{s}$, which is defined to lie on
the unit sphere. This can be directly related to the fluid model,
with the macroscopic spin vector given by $\mathbf{S}(\mathbf{x},t)=(\hbar/2)\int d\Omega3\mathbf{s}f(\mathbf{x},\mathbf{v},\mathbf{s},t)$,
where the integration element is $d\Omega=d^{3}vd^{2}s$\,\ and
the two-dimensional spin integration is carried out over the unit
sphere. For a more detailed description we refer to Ref.\ \cite{zamanian2010a}.
However, we point out that the spin part of the distribution function
is more general than a semi-classical treatment, such as that made
in, e.g.\ Refs.\ \cite{cowley,brodin-etal}, where the evolution equation
for the distribution function has the same structure as for a classical
magnetic dipole moment.

In order to calculate the weakly nonlinear low-frequency response
to an incoming transverse wave packet we make the ansatz \begin{align}
f(\mathbf{x},\mathbf{v},\mathbf{s},t)= & f_{0}(v^{2},\theta_{s})+f_{\mathrm{lf}}(z,t,\mathbf{v,}\theta_{s})\notag\\
 & +\frac{1}{2}\left[f_{1}(z,t,\mathbf{v},\mathbf{s})e^{ikz-i\omega t}+f_{1}^{\ast}(z,t,\mathbf{v},\mathbf{s})e^{-ikz+i\omega t}\right],\end{align}
where $f_{0}$ is the background distribution, $f_{\mathrm{lf}}$
is a low-frequency part due to quadratic nonlinearities and $\tilde{f}_{1}$
is a slowly modulated high-frequency wave. The background distribution
will be taken to be of the form \begin{equation}
f_{0}=\frac{n_{0}}{2\pi^{3/2}v_{T}^{3}}e^{-v^{2}/v_{T}^{2}}\left[1+\tanh\left(\frac{\mu B_{0}}{k_{B}T}\right)\cos\theta_{s}\right],\label{linear}\end{equation}
where $n_{0}$ is the equilibrium density, the thermal velocity $v_{T}$
is defined as $v_{T}=\sqrt{2k_{B}T/m}$. This is the generalization
of thermal equilibrium for the scalar distribution function, applicable
for low or moderate densities, which contains the quantum effects
on the angular spin distribution. The fully quantum mechanical background
distribution (applicable in the regime of high densities and/or strong
magnetic fields) is presented in Ref.\ \cite{zamanian2010a}. The
spin part of the distribution function above can also be written as
$f_{\uparrow}(1+\cos\theta_{s})+f_{\downarrow}(1-\cos\theta)$, where
$f_{\uparrow}\propto\exp\left(-\mu_{B}B_{0}/k_{B}T\right)$ and $f_{\downarrow}\propto\exp\left(\mu_{B}B_{0}/k_{B}T\right)$.
Thus, we see that the part proportional to $\cos\theta_{s}$ scales
as $(f_{\downarrow}-f_{\downarrow})/(f_{\uparrow}+f_{\downarrow})$,
which gives the factor $\tanh\left(\mu_{B}B_{0}/k_{B}T\right)$ in
the second term of Eq. (\ref{linear}). It is then more clearly seen
that the distribution function is separated into two parts, one for
each spin population. This can be considered as a basis for derivation
of a two-fluid model \cite{Zamanian2010b}, but we will not pursue
this further here. Note also that a semi classical reasoning would
suggest us to take the distribution function as $f_{0}\propto\exp[-E/(k_{B}T)]$,
where $E=mv^{2}/2-\mu B_{0}\cos\theta_{s}$, which differs from the
quantum mechanical angular distribution used here. We will assume
that the temperature is sufficiently low for $v_{T}\ll\omega/k$ such
that we can neglect thermal effects in the final result.

The ansatz for the incoming high-frequency field in Eq.\ \eqref{Ansatz-1}
is the same as before. The aim is then to find an equation for the
low-frequency part of the distribution function. From such an equation
we can then calculate the low-frequency response in the current density
and magnetization, and compare with the results from the previous
section. The high-frequency perturbation of the distribution function
is $f_{1}=f_{+}$ ($f_{-})$ for left-hand (right-hand) circularly
polarized waves. The expressions for $f_{\pm}$ found from Eq.\ \eqref{k1}
to linear order can be easily computed from Eq. (9) in Ref.\ \cite{lundin2010} 
\begin{eqnarray}
f_{\pm} & = & \frac{(-i)e^{\mp i\varphi_{v}}}{\omega-kv_{z}\mp\omega_{c}}\frac{q}{2m}E_{\pm}\partial_{v_{\perp}}f_{0}\nonumber \\
 &  & +\frac{e^{\mp i\varphi_{s}}}{\omega-kv_{z}\mp\omega_{cg}}\frac{\mu}{2m}\label{fpm}\\
 &  & \times\left[kB_{\pm}\left(\sin\theta_{s}\partial_{v_{z}}f_{0}+\cos\theta_{s}\partial_{\theta_{s}}\partial_{v_{z}}f_{0}\right)\pm\frac{2m}{\hbar}B_{\pm}\partial_{\theta_{s}}f_{0}\right].\nonumber \end{eqnarray}
Next, allowing for slow modulations and solving the equation to first
order in $\partial_{z}/k$, $\partial_{t}/\omega$, we note that the
zero-order solution applies after making the substitution $\omega\rightarrow\omega+i\partial_{t}$
and $k\rightarrow k-i\partial_{z}$ in Eq.\ \eqref{linear}, and
then expanding to first order in the slow derivatives. Inserting the
ansatz above into the evolution equation and consider the slow-time
scale and keeping only up to quadratic nonlinearities we obtain the
equation 
\begin{eqnarray}
 &  & \left(\partial_{t}+v_{z}\partial_{z}\right)f_{lf}+\frac{q}{m}E_{zlf}\partial_{v_{z}}f_{0}\nonumber \\
 &  & =-\left[\frac{q}{4m}\left(\tilde{\mathbf{E}}+\mathbf{v}\times\tilde{\mathbf{B}}\right)+\frac{\mu}{4m}\left(ik+\partial_{z}\right)\left(\mathbf{s}\cdot\tilde{\mathbf{B}}+\tilde{\mathbf{B}}\cdot\nabla_{s}\right)\hat{\mathbf{z}}\right]\cdot\nabla_{v}\tilde{f}_{1}^{*}\nonumber \\
 &  & \quad-\frac{\mu}{2\hbar}\mathbf{s}\times\tilde{\mathbf{B}}\cdot\nabla_{s}\tilde{f}_{1}^{*}+\mathrm{c.c.}\label{k8}\end{eqnarray}
Here we have also added a low-frequency electric field in the $z$-direction,
$E_{z\mathrm{lf}}$, which has $f_{\mathrm{lf}}$ as source. Equations
(\ref{fpm}) and (\ref{k8}) now constitute a basis for calculating
the nonlinear response in the current density and magnetization.

\subsection{Ponderomotive Force}

Inserting the first order solution $\tilde{f}_{1}$ in Eq.\ \eqref{k8},
multiplying by $qv_{z}$ and integrating over $d\Omega=d^{3}vd^{2}s$
we can derive an equation for the current density $J_{z}$. We will
neglect Landau damping associated with the particle resonances. Furthermore,
since we have assumed a low-temperature ($v_{T}\ll\omega/k$) we may
expand the denominators in $f_{1}$ to the lowest order in $v_{z}$.
All the integrals can then be evaluated.

As an intermediate step, after integrating Eq.\ \eqref{k8} we obtain
\begin{equation}
\partial_{t}J_{z\mathrm{lf}}+q\partial_{z}\int d\Omega v_{z}^{2}f_{\mathrm{lf}}-\frac{q^{2}n_{0}}{m}E_{z\mathrm{lf}}=\text{nonlinear terms}.\label{eq:Jlf}\end{equation}
The second term on the left-hand side of Eq.\ \eqref{eq:Jlf} is
a thermal correction to the low-temperature limit, which we will neglect.
Using Ampere's law for the first term on the left-hand side, we can
then obtain a closed equation for the time derivative of the low-frequency
electric field in terms of the incoming wave. After some algebra we
find \[
\left[\partial_{t}^{2}+\omega_{p}^{2}\right]E_{z\mathrm{lf}}=C_{\pm}+D_{\pm},\]
where $\omega_{p}=\sqrt{n_{0}q^{2}/m\epsilon_{0}}$ is the plasma
frequency and the ponderomotive source terms on the right-hand side
are given by \begin{align}
C_{\pm}= & \frac{q\omega_{p}^{2}\omega}{8mk^{2}(\omega\mp\omega_{c})}\left(\partial_{z}\mp\frac{\omega_{c}k}{\omega(\omega\mp\omega_{c})}\partial_{t}\right)\left\vert B_{\pm}\right\vert ^{2},\\
D_{\pm}= & -\frac{q\omega_{p}^{2}\hbar^{2}k^{2}}{16m^{3}(\omega\mp\omega_{cg})^{2}}\left(\partial_{z}+\frac{2k}{\omega\mp\omega_{cg}}\partial_{t}\right)\left\vert B_{\pm}\right\vert ^{2}\notag\\
 & \mp\frac{q\omega_{p}^{2}S_{0}}{4m^{2}(\omega\mp\omega_{cg})}\left(\partial_{z}-\frac{k}{\omega\mp\omega_{cg}}\partial_{t}\right)\left\vert B_{\pm}\right\vert ^{2}.\end{align}
 Note that the term $C_{\pm}$ is due to the classical ponderomotive
force stemming from the magnetic part of the Lorentz force, and $D_{\pm}$ 
is due to the spin effects. Taking into account the normalization
factor $n_{0}/q$ (which appears when an evolution equation for $E_{z\mathrm{lf}}$
is derived) we have perfect agreement with the fluid theory in the
previous section.

\subsection{Magnetization}

We can also obtain the low-frequency response in the magnetization.
This is done by multiplying Eq.\ \eqref{k8} by $3\mu_{B}s_{z}$
and integrating over the velocity and the spin. Thus, we obtain \begin{equation}
\partial_{t}M_{z\mathrm{lf}}+3\mu\partial_{z}\int d\Omega s_{z}v_{z}f_{\mathrm{lf}}=\text{nonlinear terms}.\label{Mlf}\end{equation}
In order to be able to evaluate the second term, we take the time
derivative of Eq. (\eqref{Mlf}) and once again use Eq.\ \eqref{k8}
to evaluate the resulting term $3\mu_{B}\partial_{z}\int d\Omega s_{z}v_{z}\partial_{t}f_{\mathrm{lf}}$.
We still neglect all the thermal contributions including the particle
resonances. To simplify the problem further, and compare with the
fluid result, we also neglect the zeroth order magnetization which
corresponds to neglecting the factor proportional to $S_{0}=(\hbar/2)\tanh\left(\mu B_{0}/k_{B}T\right)$
in Eq.\ \eqref{linear}. Also, to obtain an evolution equation for
$M_{z\mathrm{lf}}$, one takes the time derivative of Eq. (\ref{Mlf}),
substitute (\ref{k8}) for $\partial_{t}f_{\mathrm{lf}}$, apply Eq.
(\ref{fpm}) and keep only the low-frequency source terms (those proportional
to $\left\vert B_{\pm}\right\vert ^{2}$ or $\left\vert E_{\pm}\right\vert ^{2}$),
and carry out the integrals over $d\Omega$, omitting thermal corrections.
After lengthy calculations and evaluating the corresponding integrals
we eventually obtain an exact agreement with Eq. (\ref{magnetization}).
This is an important verification that the truncation of the moment
hierarchy used to close the fluid equations is valid in the low-temperature
limit, also when nonlinearities are present.

\section{Summary and Conclusions}

In this paper, we have calculated the ponderomotive force due to the
electron spin property in both a recently derived fluid model \cite{Zamanian2010b}
and in the kinetic model (see Eq.\ (83) in Ref.\ \cite{zamanian2010a})
that is the basis for the fluid theory. The kinetic result has been
evaluated in the low-temperature limit, in which case we obtain a
complete agreement with the fluid result. The fluid theory considered
here extends a simpler fluid model (e.g. Ref.\ \cite{brodin2010})
by including the spin-velocity correlations. As a result, the spin
part of the ponderomotive force found here has a contribution that
is additional to that recently derived in Ref.\ \cite{brodin2010}.
The previously derived spin force contained a term proportional to
the unperturbed spin $\mathbf{S}_{0}$ (the first term in \eqref{fluid-ponderomotive})
but not the second term in Eq.\ \eqref{fluid-ponderomotive} that is
independent of $\mathbf{S}_{0}$. Nevertheless, Ref.\ \cite{brodin2010}
found that the spin-ponderomotive force could be important even when
the unperturbed magnetization (and hence the net value of $\mathbf{S}_{0}$)
is zero, such as in an unmagnetized plasma. The reason was that a
two-fluid model of electrons was used, where the spin-up and spin-down
states were exposed to spin-ponderomotive forces pointing in opposite
directions, which induced a spin-polarized plasma. It is straightforward
to include such an approach also within Eqs.\ \eqref{Cont}-\eqref{sigma}.
Furthermore, within a nonlinear perturbation scheme, the division of
the electrons into spin-up- and spin-down-populations can be put on
a firm basis in the more complete kinetic theory.

However, provided the more subtle effects of spin-velocity correlations
are taken into account, our results here, as well as those in Ref.\
\cite{Zamanian2010b}, suggest that a two-fluid spin model (which
may capture certain effects of the microscopic spread in the spin-probability
distribution) is not needed. This results from the fact that the spin-velocity
tensor seems to capture the physics of spin-polarization, as well
as additional effects of the spin. On the other hand, it is still
too early to exclude the possibility that treating spin-up and -down
populations as different species captures new physics compared to
a one-fluid model. In particular, a difference between the one- and
two-fluid models may reveal itself when cubically nonlinear calculations
are carried out.

One of the main conclusions of this paper is that Eqs.\ \eqref{Cont}-\eqref{sigma}
agree completely with kinetic theory when thermal effects are negligible,
whereas the same would not be true when the spin-velocity tensor is
omitted. This comparison holds not only for the spin-ponderomotive
force, but also for the induced low-frequency magnetization, that
has been calculated both from the fluid and the kinetic theory. It
should be noted that the previous comparison between the current spin
fluid model and the spin kinetic theory was limited to linearized
theory. The previous expressions for the ponderomotive force were
shown to be significant in plasmas of relatively modest density and
also when the plasma was unmagnetized, due to the induced spin-polarization
\cite{brodin2010}. The additional term in the expression for the
spin ponderomotive force found here may be at least as important,
particularly when the wave frequency is close to $\omega_{cg}$, due
to the higher-order of the resonance in this new term. A detailed
evaluation of the effect of the spin-ponderomotive expression found
here is a project for future work.

\end{document}